# Towards Formal Verification of Federated Learning Orchestration Protocols on Satellites

Miroslav Popovic
*University of Novi Sad*
*Faculty of Technical Sciences*
Novi Sad, Serbia
miroslav.popovic@rt-rk.uns.ac.rs

Marko Popovic
*RT-RK Institute for Computer Based Systems*
Novi Sad, Serbia
marko.popovic@rt-rk.com

Miodrag Djukic
*University of Novi Sad*
*Faculty of Technical Sciences*
Novi Sad, Serbia
miodrag.djukic@rt-rk.uns.ac.rs

Ilija Basicevic
*University of Novi Sad*
*Faculty of Technical Sciences*
Novi Sad, Serbia
ilija.basicevic@rt-rk.uns.ac.rs

*Abstract*—Python Testbed for Federated Learning Algorithms (PTB-FLA) is a simple FL framework targeting smart Internet of Things in edge systems that provides both generic centralized and decentralized FL algorithms, which implement the corresponding FL orchestration protocols that were formally verified using the process algebra CSP. This approach is appropriate for systems with stationary nodes but cannot be applied to systems with moving nodes. In this paper, we use celestial mechanics to model spacecraft movement, and timed automata (TA) to formalize and verify the centralized FL orchestration protocol, in two phases. In the first phase, we created a conventional TA model to prove traditional properties, namely deadlock freeness and termination. In the second phase, we created a stochastic TA model to prove timing correctness and to estimate termination probability.

## I. Introduction

This research was conducted within the ongoing EU Horizon 2020 project TaRDIS [1], which aims to create a toolbox for easy programming of a broad range of distributed swarm applications, from smart grids, homes, and cities to robotics in Industry 4.0, and to navigation of LEO satellite constellations. These applications are built atop contemporary cloud-fog-edge continuum; to be smart they use some ML/AI algorithms and ideally, they should be correct by construction.

Python Testbed for Federated Learning Algorithms (PTB-FLA) [2] is a simple FL framework targeting smart IoTs in edge systems that provides both generic centralized and decentralized FL algorithms, which implement the corresponding FL orchestration protocols. To aid easy programming in tune with low-code/no-code initiative, PTB-FLA offers the 4-phase development paradigm [3] and a simple API, which is amenable both to nonprofessional developers and LLMs such as ChatGPT [4].

Recently, the PTB-FLA FL orchestration protocols were formally verified in the 2-phases process by using the process algebra Communicating Sequential Processes (CSP) and the model checker PAT [5]. The main limitation of [5] is that approach it takes is suitable for systems with stationary nodes, but cannot be applied to systems with moving nodes, such as constellations of spacecrafts, where physical timing needs to be considered, which is exactly the main motivation for this paper.

In this paper, we use celestial mechanics [6] to model spacecraft movement, and timed automata (TA) and accompanying tool UPPAAL to formalize and verify the centralized FL (CFL) orchestration protocol, in two phases. In the first phase, we created a conventional TA model to prove traditional properties, namely deadlock freeness and termination. In the second phase, we created a stochastic TA model to prove the timing correctness (the alignment of spacecraft movement and communication) and to estimate termination probability.

The main contributions of this paper are: (1) the model of a spacecraft movement in the form of the differential equation for the spacecraft's true anomaly (Section 2) that directly follows from Kepler's laws, (2) the conventional TA model of the CFL orchestration protocol, and (3) the stochastic TA model of the CFL orchestration protocol. To the best of our knowledge, this is the first paper that uses stochastic TA for formal verification of CFL orchestration protocols on constellations of spacecrafts. Unfortunately, this originality comes at the expense of closely related work of other authors, which despite our best efforts we were not able to find.

The paper is organized as follows. Section II presents the model of a spacecraft movement. Section III.1 presents the conventional TA model of the CFL orchestration protocol. Section III.2 presents the stochastic TA model of the CFL orchestration protocol. Section IV presents the more broadly related work that readers might find useful. Section V concludes the paper.

## II. Model of a Spacecraft Movement

In this section, we briefly recall the first two Kepler's laws (based on their presentation in [6]), and then we derive the differential equation for the spacecraft's true anomaly, which we used as the model of a spacecraft movement in Section III.2.

To make the exposition more concrete, let's assume that the spacecraft is flying along the elliptic path where the Earth is in its focus, see Fig. 1 (the spacecraft is in the point M and the Earth is in the point E). Let $a$ and $b$ denote the semi-major axis and the semi-minor axis, respectively, then the eccentricity $e$ is defined as:

$$e = \frac{\sqrt{a^2+b^2}}{a} \quad (1)$$

Alternatively, when given *a* and *e*, *b* may be calculated as:

$$b = a\sqrt{1-e^2} \quad (2)$$

The semi-latus rectum *p* is defined as:

$$p = \frac{b^2}{a} \quad (3)$$

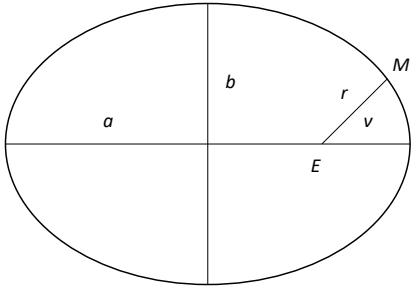

Fig. 1. The spacecraft's elliptic path.

The radius *r* and the true anomaly $v$ are related by the first Kepler's law:

$$r = \frac{p}{1 + e \cos v} \quad (4)$$

The second Kepler's law may be expressed by the following equation, where *C* is a constant:

$$r^2 \frac{dv}{dt} = C \quad (5)$$

Let *T* be the spacecraft's revolution period, then *C* is defined as:

$$C = \frac{2\pi ab}{T} \quad (6)$$

Using the following simple manipulations, we derived the equation that expresses the derivative of $v$ (also denoted as $v'$) as a function of $v$:

$$\frac{dv}{dt} = \frac{2\pi ab}{T} \frac{1}{r^2} = \frac{2\pi aba^2}{Tb^4}(1+e\cos v)^2 \quad (7)$$

The result of the derivation in (7) is the model of spacecraft movement (that is later used in Section III.2):

$$\frac{dv}{dt} = \frac{2\pi a^3}{Tb^3}(1+e\cos v)^2 \quad (8)$$

III. FORMAL VERIFICATION OF CFL PROTOCOL

The next sections present the formal verification of the CFL orchestration protocol based on the conventional and stochastic TA models, respectively.

*A. Conventional TA Model*

Fig. 2 presents the conventional TA model of the CFL orchestration protocol (the narrative of the protocol itself is available in [2]). The UPPAAL tool verified the deadlock freeness of the protocol by checking the query "A[] not deadlock" and the termination (liveness) of the protocol by checking the query "A<> (terminated == 1)".

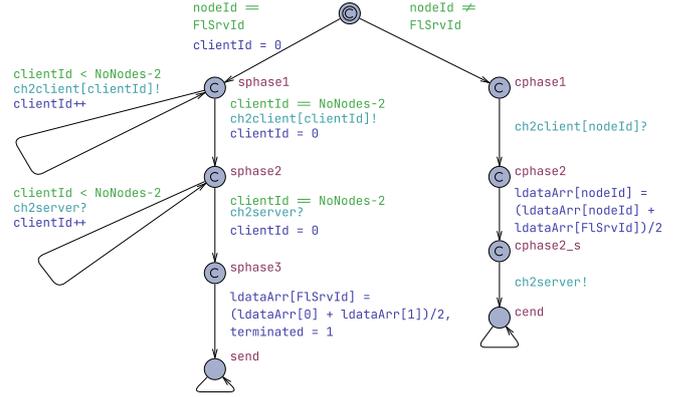

Fig. 2. The conventional TA model of the CFL orchestration protocol.

*B. Stochastic TA Model*

Let's assume that the CFL server instance resides in the ground station, whereas the two CFL client instances reside in two spacecrafts which fly on the same orbit (*a* = 10, *e* = 0.2, and *T* = 1) with the initial true anomaly $v$ of $\pi$ and 0, respectively.

The templet SV that models the spacecraft instance *nodeId* on the orbit *orbitId* is shown in Fig. 3, where *nus*[*nodeId*] stores the value of $v$ for this instance, the location *loop* invariant "*nus*[*nodeId*]' == nup(*nodeId*, *orbId*)" encodes the differential equation (8), and the function nup returns the new value of $v$ according to (8). Initially, $v$ is set to a value in range [0, 2$\pi$), and once it reaches 2$\pi$, this instance broadcasts the signal "*reset*[*nodeId*]", which means that this spacecraft reached the periapsis, and that the communication between the ground station and this spacecraft should take place.

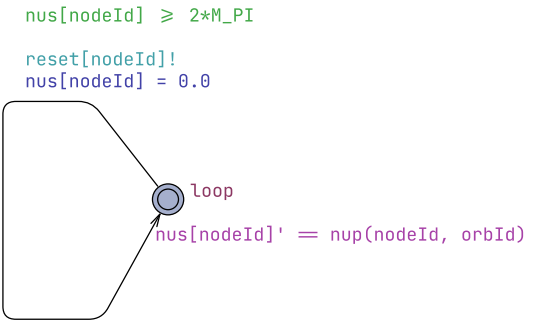

Fig. 3. The stochastic TA model of the spacecraft.

Fig. 4 presents the stochastic TA model of the CFL orchestration protocol, which is implemented as the templet FlCentralized that models the CFL instance *nodeId*. The server instance (*nodeId* == FlSrvid) takes the left branch whereas the client instance takes the right branch.

The first CFL phase takes place during the first revolution of the spacecrafts: when the server receives for the signal *reset*[*clientId*] in the location *sphase1_t*, it sends the signal *ch2client*[*clientId*] to the client, which in turn receives that signal and proceeds to the location *cphase2_t*. After sending the second signal, the server goes to the location *sphase2*.

The second CFL phase takes place during the second revolution of the spacecrafts: when the client *nodeId* (== *clientId*) receives the signal *reset*[*nodeId*] in the location *cphase2_t*, it sends the signal *ch2server* to the server, which in turn receives that signal; the client goes to the location *cend*. After receiving the second signal, the server goes to the location *sphase3*, and from there to the location *send*. At that point CFL is terminated.

We used the query "*Pr*[<=3](<> *server.send*)" to estimate the probability of the CFL orchestration protocol termination (which happens when the CFL instance *server* reaches the location *send*) before 3 time-units. UPPAAL tool estimated

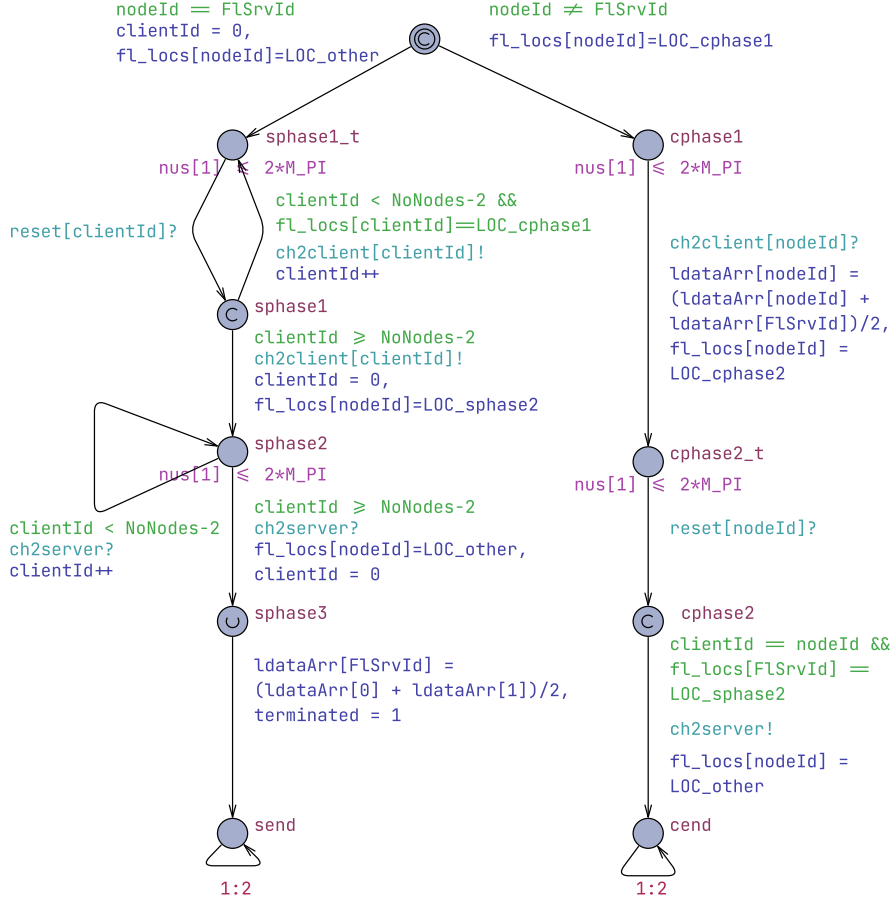

Fig. 4. The stochastic TA model of the CFL orchestration protocol.

Recall that initially *nus*[1] is set to 0, therefore the invariant "nus[1] ≤ 2*M_PI" of the locations *sphase1_t* and *cphase2_t* correspond to the first and the second revolution of the second spacecraft, respectively. Since *nus*[0] is initially $\pi$, the first spacecraft always flies through the periapsis before the second spacecraft.

To check the alignment of spacecrafts' movement and their communication with the ground station we created a simulation query that traces both true anomalies and current locations of the CFL server and the CFL clients. The simulation diagram plot by UPPAAL looks like a Gantt chart drawn atop a timing of analog signals. The former comprises the three staircase-like lines showing location changes, see Fig. 5.

Obviously, the alignment is perfect, see how: (1) the first client (green line) reaches the location *cphase2_t* at the time point $t = T/2$, and the location *cend* at the time point $t = 3T/2$, (2) the second client (blue line) reaches the location *cphase2_t* at the time point $t = T$, and the location *cend* at the time point $t = 2T$, and (3) the server (pink line) reaches the location *sphase2* at the time point $t = T$, and the location *send* at the time point $t = 2T$. Both *nus*[0] and *nus*[1] have the period T, and they start from $\pi$ and 0 (see the ordinate at the time point $t = 0$), respectively.

the CFL termination probability as "(72/72 runs) *Pr*(<> …) in [0.950056,1] (95% CI)", which means that the 95% confidence interval [0.950056,1] was obtained in 72 runs.

IV. RELATED WORK

UPPAAL is a well-known tool and there is a vast literature related to it. In this paper, we mostly used the paper [7] to construct the stochastic TA model in Section III.2.

The stochastic TA model has two main limitations: (i) it encompasses only two spacecrafts on a single orbit, and (ii) it uses a trivial FL algorithm. In our future work, we plan to model: (i) more spacecrafts on more orbits e.g., by using data from [8], and (ii) some realistic algorithms e.g., from [9].

The following three papers are more broadly related and orthogonal to this paper. The contributions of the paper [10] are the introduction of concurrency considerations at the early space mission design phases and the use of UPPAAL for the mission feasibility check.

Similarly, the paper [11] is a comparative scalability study of various model-checking algorithms using an arbitrarily scalable operational design describing the mode management of a satellite. While [10] and [11] consider concurrency and model checking for a single spacecraft, this paper considers communication among the ground station and more spacecrafts.

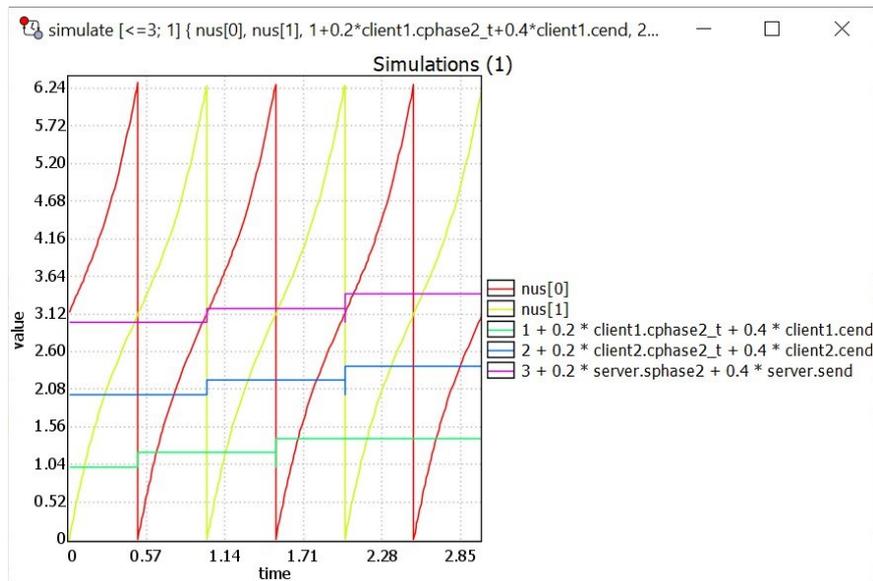

Fig. 5. The UPPAAL tool simulation diagram.

In the paper [12], a TDMA-centric inter-satellite communication specification is represented and verified using PVS (Prototype Verification System). A theorem proving approach was followed to check well-formedness (no contradiction) of the specification and verify the requirements. While [12] only mentions the function distance that returns the distance between the two given satellites within the given timeslot, but does not provide its implementation, this paper provides the way towards modeling the motion of more satellites, that would enable calculating distances between the satellites.

## V. Conclusion

The main contributions of this paper are: (1) the model of a spacecraft movement in the form of the differential equation for the spacecraft's true anomaly, (2) the conventional TA model of the CFL orchestration protocol, and (3) the stochastic TA model of the CFL orchestration protocol.

The main advantage of the stochastic TA model over the conventional is that it may be used to prove timing correctness i.e., alignment of spacecraft movement and communication, coupled with the estimated termination probability in 95% confidence interval [0.95, 1].

The main limitations of the stochastic TA model are that it models (i) only two spacecrafts on a single orbit and (ii) a trivial FL algorithm. In the future work, we plan to extend it to: (i) more spacecrafts and orbits and (ii) realistic algorithms.

### Acknowledgment

Funded by the European Union (TaRDIS, 101093006). Views and opinions expressed are however those of the author(s) only and do not necessarily reflect those of the European Union. Neither the European Union nor the granting authority can be held responsible for them.